\documentstyle[editedvolume,numreferences]{crckapb} %

\def\12{\frac{1}{2}} 
\def\tk{\tilde{k}} 
\def\tg{\tilde{g}} 
\def\tH{\tilde{H}} 
\def\tA{\tilde{A}}

\begin{opening}
\title{D-BRANE ACTIONS, INTRINSIC  GEOMETRY AND DUALITY}
\author{M.\ ABOU ZEID} 
\institute{Physics Department, Queen Mary and Westfield College, \\ 
Mile End Road, London E1 4NS, U.\ K.\ } 
\end{opening}

\begin{document}

\begin{abstract}  
We discuss an alternative form of the   
supersymmetric D-$p$-brane  
action which is quadratic in derivatives of $X$ and linear in  
$F_{\mu \nu}$. This action involves an auxiliary worldvolume 
tensor and generalises the  
simplification of the Nambu-Goto action for $p$-branes  
using  a symmetric metric. When the 
worldvolume gauge  field is abelian, it 
appears as a Lagrange multiplier, and solving the constraint 
gives the dual form of the ($p+1$)-dimensional action  
with a $p-2$ form gauge field instead of a vector gauge field. This is illustrated by the example of the dual D-2-brane action, for which the known result is recovered. 
\end{abstract}

\section{Supersymmetric D-$p$-Brane Actions}

The effective action for superstring theory\index{effective action} in the presence of a D-$p$-brane
is the sum $S=S_D +S_{(p)}$ of the bulk supergravity 
action $S_D$\index{supergravity action} and the effective world-volume action $S_{(p)}$\index{effective world-volume action}.
The latter is the sum of a Dirac-Born-Infeld\index{D-brane action!Dirac-Born-Infeld term} type action~\cite{RGL,FT} 
and a Wess-Zumino\index{D-brane action!Wess-Zumino term}
term~\cite{ML,MD}, and it was shown in 
refs.~\cite{JSetal1,JSetal2,CGetal,BT}, to 
be  invariant
under a local fermionic  symmetry\index{local fermionic symmetry} which reflects the
fact that the D-brane  is BPS-satured and breaks half the space-time 
supersymmetries~\cite{JP}. 

We begin with a brief review of the usual form of $S_{(p)}$. The 
(flat) superspace\index{superspace} coordinates are the $D=10$ space-time coordinates
$X^i$ and the Grassmann coordinates $\theta$, which are space-time spinors
and world-volume scalars. For the IIa superstring (even $p$), $\theta$ is
Majorana but not Weyl while in the IIb superstring there are two
Majorana-Weyl spinors $\theta_\alpha$ ($\alpha =1,2$) of the same chirality.
The superspace (global) supersymmetry transformations\index{supersymmetry transformations} are
\begin{equation} 
\delta_\epsilon \theta = \epsilon \ \ , \ \ \delta_\epsilon X^i = 
\overline{\epsilon} \Gamma^i \theta .
\end{equation}                          
The world-volume theory has global type IIa or type IIb super-Poincar\'{e}
symmetry\index{super-Poincar\'{e} symmetry} and is constructed using the supersymmetric one-forms $\partial_\mu
\theta$ and
\begin{equation} 
\Pi_{\mu}^{i} = \partial_\mu X^i -\overline{\theta} \Gamma^i \partial_\mu 
\theta .
\end{equation} 
The induced world-volume metric\index{induced metric} is
\begin{equation} 
G_{\mu \nu} = G_{ij} \Pi_{\mu}^{i} \Pi_{\nu}^{j} .
\end{equation} 
The supersymmetric world-volume gauge field-strength\index{world-volume gauge field} $\cal F$ is defined by
${\cal F}=F-B$, and a convenient choice for 
the two-form $B$ is~\cite{PKT} 
\begin{equation} 
B=-\overline{\theta} \Gamma_{11} \Gamma_i d\theta ( dX^i +\12 \overline{\theta} 
\Gamma^i d\theta )
\label{Bchoice} 
\end{equation} 
when $p$ is even or the same formula with $\Gamma_{11}$ replaced with
the Pauli matrix $\tau_3$ when $p$ is odd. With the choice~(\ref{Bchoice}), 
$\delta_\epsilon
B$ is an exact two-form and $\cal F$ is supersymmetric for an appropriate
choice of $\delta_\epsilon A$~\cite{PKT,JSetal2}.

The leading contribution to the effective supersymmetric D-$p$-brane action\index{supersymmetric D-brane action} 
with constant dilaton\index{dilaton} and zero cosmological constant\index{cosmological constant} then takes the form
\begin{equation} 
S_{(p)}  =  -T_p \int d^n \sigma e^{-\phi} \sqrt{-\det ( G_{\mu \nu} +  
{\cal F}_{\mu \nu})} + T_p \int_{W_{p+1}}C e^{\cal F} 
\label{action}
\end{equation} 
The constant $T_p$ is the $p$-volume tension\index{volume tension} with mass dimension
$p+1$, and $W_{p+1}$ denotes the $(p+1)$-dimensional worldvolume of the brane. Here $C$ represents the complex
of differential forms $C=\sum^{9}_{r=0}C^{(r)}$ where the $C^{(r)}$ are the
pull-backs of superspace forms $C^{(r)}=d\overline{\theta} T^{(r-2)}
d\theta$ for certain $r-2$ forms $T^{(r-2)}$ given explicitly 
in~\cite{CGetal,BT,JSetal1,JSetal2}, and it is understood that the
$p+1$ form part of $Ce^{\cal F}$, which is $C^{(p+1)}+C^{(n-2)}{\cal F}
+\frac{1}{2}C^{(n-4)} {\cal F}^2 +\ldots$, is selected. The bosonic part
of the Wess-Zumino term gives the coupling of the brane to the
background Ramond-Ramond $r$-form gauge fields (where $r$ is odd for type
IIa and even for type IIb). The case of the 9-form potential is special because its field equation forces the dual of its field strength to be a constant $m$. The backgrounds with $m=0$ are the familiar type IIa ones, while backgrounds
with $m\neq 0$ are those of massive type IIa supergravity~\cite{R}. In the
$m\neq 0$ case the  Wess-Zumino term given in~(\ref{action}) requires
$m$-dependent Chern-Simons modifications, and the bosonic part of these were derived in ~\cite{GHT}. The constant $m$ will be taken to be zero here, so that the 9-form potential vanishes; the generalisation of the discussion below to the more general situation will be given elsewhere.

The action~(\ref{action}) is supersymmetric and invariant under local kappa
symmetry~\cite{CGetal,BT,JSetal1,JSetal2}\index{kappa symmetry}. 
However, just as in the case of the usual Nambu-Goto action for a $p$-brane,
the non-linearity of~(\ref{action}) is inconvenient for many purposes. In
particular, dualising the action has proved difficult in this approach,
and has only been achieved for $p<5$~\cite{PKT,CS,AS,AAT,YL,JSetal3}. It is therefore
useful to generalise the simplification of the Nambu-Goto action\index{Nambu-Goto action} using an
auxiliary world-volume metric\index{auxiliary world-volume metric} $g_{\mu \nu}$ to the case of actions 
involving world-volume gauge fields in addition to the induced metric $G_{\mu
\nu}$. This was achieved in ref.~\cite{AH} by introducing an auxiliary
tensor field
\begin{equation}
k_{\mu \nu} \equiv g_{\mu \nu} +b_{\mu \nu}
\end{equation}
with both a symmetric part $g_{\mu \nu}$ and an antisymmetric part $b_{\mu 
\nu}$. The supersymmetric D-brane action which is classically equivalent to
(\ref{action}) is 
\begin{equation} 
S  =  -\12 T_p ' \int d^n \sigma e^{-\phi} \sqrt{-k} \left[ k^{\mu \nu} 
(G_{\mu \nu} +{\cal F}_{\mu \nu} ) -(p-1) \Lambda \right]  +T_p \int_{W_{n}} 
C e^{\cal F} ,
\label{newaction}
\end{equation} 
where $k\equiv \det k_{\mu \nu}$ and $\Lambda$ is a constant. The 
inverse tensor $k^{\mu \nu}$
satisfies
\begin{equation}
k^{\mu \nu}k_{\nu \rho} = \delta^\mu_\rho . 
\end{equation}
For $p\neq 1$, the $k_{\mu \nu}$ field equation implies
\begin{equation}
G_{\mu \nu} +{\cal F}_{\mu \nu} =\Lambda k_{\nu \mu}
\end{equation}
and substituting back into~(\ref{newaction}) yields the supersymmetric
D-brane action in the DBI form~(\ref{action}) , with the 
constants $T_p$ ,$T'_p$
related by
\begin{equation}
T'_p =\Lambda^{\12 (p-1)} T_p  .
\end{equation}
For $p=1$, the action~({\ref{newaction}) is invariant under the generalised
Weyl transformation\index{Weyl transformation} 
\begin{equation}
k_{\mu \nu} \rightarrow \omega (\sigma ) k_{\mu \nu}
\label{Weyl}
\end{equation}
and the $k_{\mu \nu}$ field equation implies
\begin{equation}
G_{\mu \nu} +{\cal F}_{\mu \nu} = \12 k^{\rho \sigma} (G_{\rho \sigma} +{\cal 
F}_{\rho \sigma}) k_{\nu \mu} .
\end{equation}

The first term in the action~(\ref{newaction}) is {\em linear} in 
$(G_{\mu \nu} +{\cal F}_{\mu \nu})$ and {\em quadratic} in $\partial X$, 
so it is easier to analyse than~(\ref{action}). Moreover, being classically
equivalent to~(\ref{action}), the action~(\ref{newaction}) is 
supersymmetric and invariant under local kappa symmetry; this 
can also be checked explicitly.

\section{Dual Super D-$p$-Brane Actions}

The dualisation\index{D-brane action!dualisation} of the  action~(\ref{newaction}) can be achieved
by adding a Lagrange term\index{Lagrange multiplier term} $\12 \tH^{\mu \nu} (F_{\mu \nu}-
2\partial_{[\mu}A_{\nu]})$ imposing the constraint $F=dA$. Here the 
anti-symmetric tensor density $\tH^{\mu \nu}$
is a Lagrange multiplier which can be integrated out to regain the original
action~(\ref{newaction}). Alternatively, integrating out over $A_\mu$ imposes
the constraint
\begin{equation}
\partial_\mu \tH^{\mu \nu} =0 ,
\label{constr}
\end{equation}
which can be solved in terms of a $p-2$ form $\tA$:
\begin{equation}
\tH^{\mu \nu} = \frac{1}{(p-1)!} \epsilon^{\mu \nu \rho \gamma_1 \ldots
\gamma_{p-2}} \partial_{[\rho} \tA_{\gamma_1 \ldots \gamma_{p-2}]} ,
\label{deftH}
\end{equation}
where $\epsilon^{\mu \nu \rho \ldots}$ is the alternating tensor density. Now $F$ is an auxiliary 2-form occuring algebraically; we emphasize this 
by rewriting $F\rightarrow L$, so that the action is
\begin{eqnarray}
& & -\12 T'_p \int d^{p+1} \sigma \left\{ \sqrt{-k} [ k^{\mu \nu}
(N_{\mu \nu} +L_{\mu \nu} ) -(p-1) \Lambda ] +\12 \tH^{\mu \nu}
L_{\mu \nu} \right\} \nonumber \\ & & +T_p \int_{W_{p+1}} Ce^{L-B} .
\end{eqnarray}
The field equation for $L$ is then
\begin{equation}
\sqrt{-k} k^{[\mu \nu ]} +\12 \tH^{\mu \nu} +
\frac{\delta f (L)}{\delta L_{\mu \nu}} =0 ,
\label{feL}
\end{equation}
where the potential $f(L) = Ce^{L-B}$ is a polynomial of order [$\12 (p+1)$] in
$L$ (i.~e.\ the integer part of $\12 (p+1)$), so that the field 
equation~(\ref{feL}) is of order [$\12 (p+1)-1$], which is at most quartic
for $p<9$. In particular, it is quadratic for $p<5$, so that the dual
supersymmetric actions for the corresponding D-branes should be obtainable
straigthforwardly; this will be illustrated in an example below, and 
will be discussed further elsewhere. Here we will consider only the cases
in which  the action~(\ref{newaction}) is linear in $F$.

\section{The D-Membrane}

The dualisation of (the bosonic part of) the D-string action starting from the form~(\ref{newaction}) was discussed in ref.~\cite{AH}; here we consider the case of the D-membrane of type IIa superstring theory\index{D-membrane}. For $p=2$, (\ref{newaction}) reduces to the following form of the D-2-brane action in the string metric, 
\begin{equation} 
S  =  -\12 T_2 ' \int d^3 \sigma e^{-\phi} \sqrt{-k} \left[ k^{\mu \nu}  
(G_{\mu \nu} +{\cal F}_{\mu \nu}) -\Lambda \right]  +T_2 
\int_{W_{3}} C^{(3)} +C^{(1)} {\cal F} 
\end{equation} 
with $T_2 '= \sqrt{\Lambda} T_2$. Adding a Lagrange multiplier term $\12 \tH^{\mu \nu}(F_{\mu \nu}-
2\partial_{[\mu}A_{\nu ]})$ to this action and  
integrating out the gauge field $A$ gives the constraint~(\ref{constr}), 
which is solved in $p+1=3$ dimensions by 
\begin{equation} 
\tH^{\mu \nu} = \epsilon^{\mu \nu \rho} \partial_\rho S 
\end{equation} 
for some scalar field $S$. Substituting this back in the action and using a 
density\index{density} $\tk^{\mu \nu} = \sqrt{-k} k^{\mu \nu}$ with $\tk \equiv \det 
\tk^{\mu \nu} = -\sqrt{-k}$ gives 
\begin{eqnarray} 
S  & =  & -\frac{\sqrt{\Lambda}}{2} T_2 \int d^3 \sigma \left\{ e^{-\phi} \left[ \tk^{\mu \nu} 
(N_{\mu \nu}+L_{\mu \nu}) +\Lambda \tk \right] +\epsilon^{\mu \nu \rho} 
K_\rho L_{\mu \nu} \right\}  \nonumber \\ & & +T_2 \int_{W_{3}} 
C^{(3)}-C^{(1)}B ,
\label{D2act1}
\end{eqnarray} 
where $L$ is an auxiliary two-form and we have defined 
\begin{equation} 
K_{\rho} \equiv \12 \partial_{\rho} S -\frac{1}{\sqrt{\Lambda}} C_{\rho}^{(1)} .
\end{equation} 
  
Integrating out $L$ yields the constraint
\begin{equation}
e^{-\phi}\tk^{[\mu \nu ]} +\epsilon^{\mu \nu \rho} K_\rho =0 ,
\end{equation}
so that
\begin{equation}
\tk^{\mu \nu} =\tg^{\mu \nu} +\tilde{\cal H}^{\mu \nu}
\label{solktilde}
\end{equation}
where $\tilde{\cal H}^{\mu \nu}=-\epsilon^{\mu \nu \rho} e^\phi K_\rho$, and $\tg^{\mu \nu}$ is the symmetric tensor density defined by $\tg^{\mu \nu}=\tk^{(\mu \nu)}$. Then
$\det (\tk^{\mu \nu})= \det (\tg^{\mu \nu}) \Omega$, where
$\Omega = \det (\delta^\nu{}_\rho +\tilde{\cal H}^\nu{}_\mu )$. Using
the identity $\det (1+X) =1-\12 tr X^2 $ valid for $n=3$ and $X$ antisymmetric,
we find  
\begin{eqnarray}
\Omega & = & 1-\12 \tilde{\cal H}^\nu{}_{\mu} \tilde{\cal H}_\nu{}^{\mu} 
 =   1-\12 e^{2\phi} 
\tg_{\mu \rho} \tg_{\nu \sigma} 
\epsilon^{\rho \nu \kappa} \epsilon^{\sigma \mu \delta} K_\kappa K_\delta 
 \nonumber \\ & = & 1+\12 e^{2\phi} g^{\mu \nu}K_\mu K_\nu  ,
\label{crux}
\end{eqnarray}
where $g^{\mu \nu} =
\frac{1}{\sqrt{-g}} \tg^{\mu \nu}$ is  a symmetric tensor with inverse $g_{\mu \nu}$, and $g\equiv \det(g_{\mu \nu})$. Using these results, the first term in~(\ref{D2act1}) becomes 
\begin{equation}
S  =  -\frac{\sqrt{\Lambda}}{2} T_2 \int d^3 \sigma  e^{-\phi} \sqrt{-g} \left[ g^{\mu \nu}
G_{\mu \nu} -\Lambda \Omega \right]  +\frac{\sqrt{\Lambda}}{2}
T_2 \int_{W_3} B \wedge K .
\label{presque}
\end{equation}
The field equation for the metric $g_{\mu \nu}$ implies
\begin{equation}
 g_{\mu \nu} = 
\frac{1}{\Lambda} \left( G_{\mu \nu} -\frac{\Lambda}{2} e^{2\phi} K_\mu K_\nu \right)  .
\end{equation}
Substituting this back in~(\ref{presque}) and adding the second term in~(\ref{D2act1}) yields the dual action
\begin{eqnarray}
S  & = &  -T_2 \int d^3 \sigma e^{-\phi} \sqrt{ -\det \left[ G_{\mu \nu}
-\frac{\Lambda}{8} e^{2\phi} ( \partial_\mu S -\frac{2}{\sqrt{\Lambda}}C_{\mu}^{(1)}
) ( \partial_\nu S -\frac{2}{\sqrt{\Lambda}}C_{\nu}^{(1)}
) \right] } \nonumber \\ & & +   T_2 \int_{W_3} \left( -C^{(3)} +\frac{\sqrt{\Lambda}}{2}
B dS \right)
\end{eqnarray}
which agrees with that of~\cite{PKT,CS,AAT,YL,JSetal3} if we set $\Lambda =4$. The dilaton dependence can be removed by a rescaling of the superspace coordinates, and this yields the standard M theory membrane action\index{M theory membrane} with a circular dimension, as expected on the basis of the relationship between type IIa superstring theory\index{IIa superstring theory}
and M theory\index{M theory}.

\vspace{.2cm}

{\bf Acknowledgements} 
\\  
I would like to thank Christopher Hull for collaboration and  
suggestions, and Vijay Balasubramanian, Emil Martinec, Dimitri 
Sorokin and Paul Townsend for helpful 
discussions. I am also grateful to the organisers of the school for their 
support.

\end{document}